\documentclass[reprint,showpacs,groupedaddress,footinbib,citeautoscript]{revtex4-1}

\usepackage{graphicx}
\usepackage{amsmath}
\usepackage{amssymb}
\usepackage{mathrsfs}
\usepackage{gensymb}
\usepackage{bm}
\usepackage{array}
\usepackage{natbib}

\newcolumntype {s}[1]{@{\hspace{#1}}} 
\usepackage{color}
\usepackage[usenames,dvipsnames]{xcolor}
\usepackage{wrapfig}
\usepackage[normalem]{ulem}

\begin{document}

\title{Optical Spectroscopy of SmN: Evidence for 4$f$ Transport}

\author{W.F.~Holmes-Hewett}
\affiliation{The MacDiarmid Institute
for Advanced Materials and Nanotechnology and The School of Chemical and Physical Sciences, Victoria University of Wellington,
PO Box 600, Wellington 6140, New Zealand}

\author{R.G.~Buckley}
\affiliation{The MacDiarmid Institute
for Advanced Materials and Nanotechnology and Robinson Research Institute, Victoria University of Wellington,
PO Box 600, Wellington 6140, New Zealand}

\author{B.J.~Ruck}
\affiliation{The MacDiarmid Institute
for Advanced Materials and Nanotechnology and The School of Chemical and Physical Sciences, Victoria University of Wellington,
PO Box 600, Wellington 6140, New Zealand}

\author{F.~Natali}
\affiliation{The MacDiarmid Institute
for Advanced Materials and Nanotechnology and The School of Chemical and Physical Sciences, Victoria University of Wellington,
PO Box 600, Wellington 6140, New Zealand}

\author{H.J.~Trodahl}
\affiliation{The MacDiarmid Institute
for Advanced Materials and Nanotechnology and The School of Chemical and Physical Sciences, Victoria University of Wellington,
PO Box 600, Wellington 6140, New Zealand}

\date{\today}

\pacs{71.27.+a,		
	 75.50.Pp	
          }

\begin{abstract}

The rare-earth nitride ferromagnetic semiconductors owe their varying magnetic properties to the progressive filling of 4$f$ shell across the series. Recent electrical transport measurements on samarium nitride, including the observation of superconductivity, have been understood in terms of a contribution from a 4$f$ transport channel. Band structure calculations generally locate an empty majority 4$f$-band within the conduction band although over a wide range of possible energies. Here we report optical reflection and transmission measurements on samarium nitride between 0.01 eV to 4 eV, that demonstrate clearly that the 4$f$ band forms the bottom of the conduction band. Results at the lowest energies show no free carrier absorption, indicating a semiconducting ground state, and supporting earlier conclusions based on transport measurements.

\end{abstract}

\maketitle

\section{Introduction}

The remarkable behaviours precipitated by electron correlations in condensed matter systems determines that they have developed an intense interest within the solid-state physics community \cite{Coleman2007,Degiorgi1999,Adler2019}. Within that literature the most striking, the most potentially exploitable, behaviour occurs when a very narrow, high-mass, strongly localised band with a large density of states crosses and hybridises with a conventional conduction band \cite{Riseborough2000,Dzero2016}. There is, however, a paucity of simple systems displaying strongly correlated electron characteristics. Within this picture the LN series (L a lanthanide element) show promise as a valuable testing ground for simply structured strongly correlated materials due to their rock-salt structure and varying occupancy of the 4$f$ shell along the series \cite{Larson2007,Johannes2005,Aerts2004,Adler2019}.
Here we report optical spectroscopy which shows that a localised band forms the conduction band minimum in the ferromagnetic semiconductor SmN.


The members of the LN series share the simple NaCl crystal structure with 4$f$ bands crossing the 2$p$ valence band (VB) or the 5$d$ conduction band (CB) at an energy that varies among the fourteen members. Most show semiconducting behaviour amenable to tuning the Fermi energy into the heavy-mass 4$f$ bands with modest doping levels \cite{Natali2013,Larson2007}. They have widely ranging magnetic behaviours associated with the varying 4$f$ configuration, affecting both the quantum state and its interaction with the electron transport bands, and in at least one case, SmN, a superconducting state is found to occur below 4~K\cite{Anton2016a}. 

The prototypical LN is GdN \cite{Granville2006}, which serves as a valuable benchmark to compare with the other members of the series, based in part on the simplicity offered by the half-filled 4$f$ shell in the Gd$^{3+}$ ion. GdN shows ferromagnetic alignment below its Curie temperature of $\sim$~70~K \cite{Plank2011} and a fully spin polarised conduction channel up to doping levels of $10^{21}$cm$^{-3}$ \cite{Trodahl2017a}. The ${}^{8}S_{7/2}$ configuration of its 4$f$ shell features $\mathrm{L}=0$, $\mathrm{S}=\mathrm{J}=7/2$ ensuring that the majority- (minority-) spin 4$f$ band lies 7~eV below (5~eV above) the band gap \cite{Larson2007,Trodahl2007}. This in turn ensures no strong electron correlation characteristics with the N2$p$ VB and Gd5$d$ CB both of which are well separated from the 4$f$ bands. GdN features a VB maximum at $\Gamma$ and CB minimum at X, as is expected also for the remaining LN. 


\begin{figure}
\centering
\includegraphics[width=\linewidth]{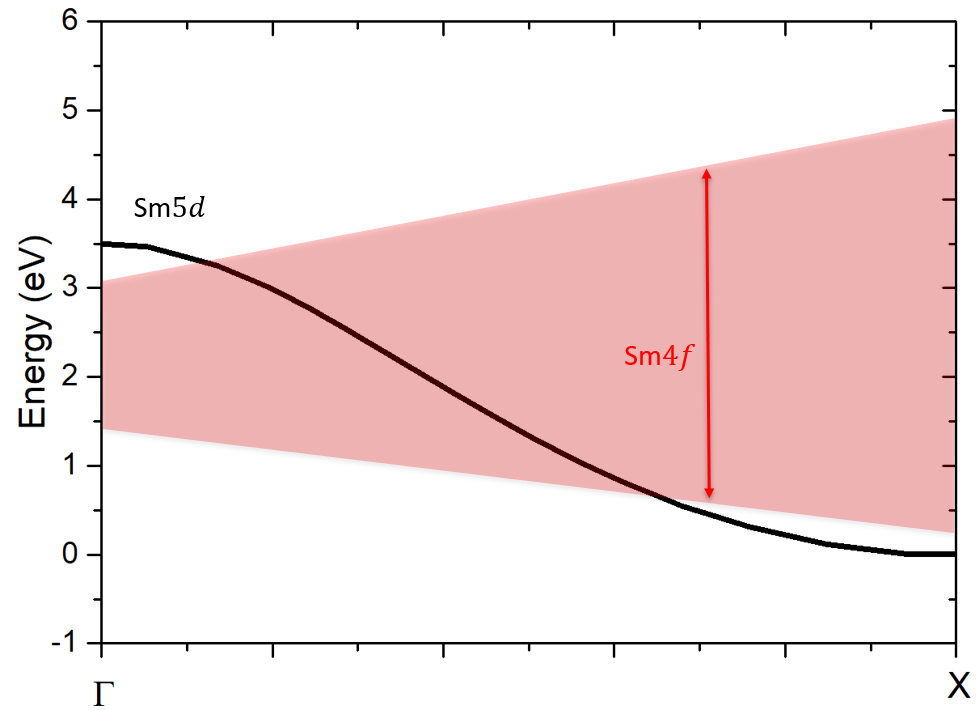}
\caption{(Colour online) Schematic band structure diagram showing the range of calculated locations of lowest unoccupied Sm4$f$ majority spin band (shaded), along with the 5$d$ band of SmN for reference (solid black). Plots are scaled such that zero energy is taken at the CB minimum. Data reproduced from Refs[\onlinecite{Larson2007,Cheiwchanchamnangij2014,Som2017,Morari2015a}].}
\label{fig:BS-4f}
\end{figure}

With two fewer electrons, SmN adopts a ${}^6H_{5/2}$ configuration leaving two unfilled majority spin 4$f$ states. A free Sm$^{3+}$ ion has a Land$\mathrm{\acute{e}}$ $g$ factor of $g_J=2/7$, and a resulting ground state moment of $\mu_B g_J J=0.71~\mu_B$. In the crystal field the paramagnetic state of SmN carries a magnetic moment of 0.45~$\mu_B$, falling to 0.035~$\mu_B$ under the combined crystal field and exchange interaction below the $\sim$~30~K Curie temperature \cite{Meyer2008,Mcnulty2016}. Such a small magnetic moment results form a nearly complete cancellation between opposing spin and orbital contributions in the partially filled 4$f$ shell. 

The electronic structure of SmN has been investigated within various density functional theory treatments which show the lowest unoccupied majority-spin 4$f$ band variously \cite{Larson2007,Cheiwchanchamnangij2014,Som2017,Morari2015a} to lie from the bottom of the 5$d$ conduction band to some few eV higher. This is shown in Figure~\ref{fig:BS-4f}, along with the 5$d$ band which forms the CB minimum across many of the LN series. Interestingly none of the predictions place the 4$f$ band below the 5$d$, with even the lowest predicted energy leading only to a strong 4$f$/5$d$ hybridisation near the bottom of the 5$d$ band which then repels the main weight of the 5$d$ states to higher energy and leaves the potential of a hybridisation gap above the 4$f$ band. Recent experimental reports of an enhanced anomalous Hall effect \cite{Holmes-Hewett2018}, and superconductivity \cite{Anton2016a} in SmN both show some degree of 4$f$ influence on the conduction channel, implying that the 4$f$ band does indeed exist near, if not form, the CB minimum. Nearly all calculations also result in a semi-metallic ground state, in stark contrast to recent experimental reports all of which support the semiconducting conclusion \cite{Natali2013,Meyer2008,Anton2013,Anton2016a,Holmes-Hewett2018}. It is clearly important to establish a band structure of SmN and determine the role played by the 4$f$ band.

In the present manuscript we inform the calculations with 0.01~eV - 4~eV optical spectroscopy delineating these features. We compare measurements on SmN and GdN in the context of their electronic structures and locate the 4$f$ band, which we show forms the bottom of the conduction band in SmN.



\section{Experimental Methods}

Thin films of SmN and GdN were grown as detailed in our recent review \cite{Natali2013}. Various substrates, suited to each optical or characterisation measurement, were included in each growth to ensure consistency across the various measurements. Films were capped with $\sim$~100~nm of AlN for protection from the damaging effects of atmospheric oxygen and water vapour. Electrical measurements were conducted on sapphire [0001] 10$\times$10$\times$0.5~mm substrates with pre-deposited Au contacts in a van der Pauw configuration. These measurements, conducted in an Oxford Cryogenics close cycle cryostat between 300-4~K, showed resistivities increasing rapidly at low temperatures for both SmN and GdN implying low carrier densities and in turn a low concentration of nitrogen vacancies. Reflection and transmission measurements were conducted using a Brooker Vertex 80v Fourier transform spectrometer in which samples  were mounted on a cold finger inside an Oxford Cryogenics flow though LHe cryostat. Low energy measurements (0.01~eV-1.2~eV) were conducted on Si substrates where the material is largely transparent, higher energy measurements (0.12~eV to 4~eV) were conducted on sapphire substrates for the same reason. Measurements were conducted as a function of temperature, on both substrate materials down to cryogenic temperatures well below the recognised Curie temperatures of SmN (30~K) and GdN (70~K). Reflection measurements were taken using a 250~nm Al film as reference and data from Ehrenreich \cite{Ehrenreich1963} to correct the resulting spectrum.  

The software package RefFit \cite{RefFit}, which uses an inherently Kramers-Kronig consistent sum of Lorentzians to represent the dielectric function $\epsilon(\omega)$ of a material, was used to simultaneously reproduce reflection and transmission measurements on both substrate materials. Analysis for any measurement was conducted only over the range of the measurement (e.g. 0.12~eV to 4~eV for sapphire). To account for all high energy behaviour above the measurement range a constant value $\epsilon_\infty$ was first determined by matching the amplitude and phase of the interference fringes in the reflectivity data. Reflection and transmission spectra were reproduced independently for the substrates and capping layers enabling the contribution of the LN layers to be investigated independently. When reproducing the LN layers, Lorentzians with a width of 100~cm$^{-1}$ were used, these were spaced every 100~cm$^{-1}$ from 300~cm$^{-1}$ to 40,000~cm$^{-1}$. Phonon absorption near 250~cm$^{-1}$ was reproduced using a single Lorentzian. Finally a zero frequency Drude term was used with parameters constrained to the measured DC resistivity. The resulting dielectric functions were then used to calculate the real part of the optical conductivity $\sigma_1(\omega)$, which is now free from interference and absorption effects from the substrate and capping layer.

\section{Results and discussion}

\begin{figure}
\centering
\includegraphics[width=\linewidth]{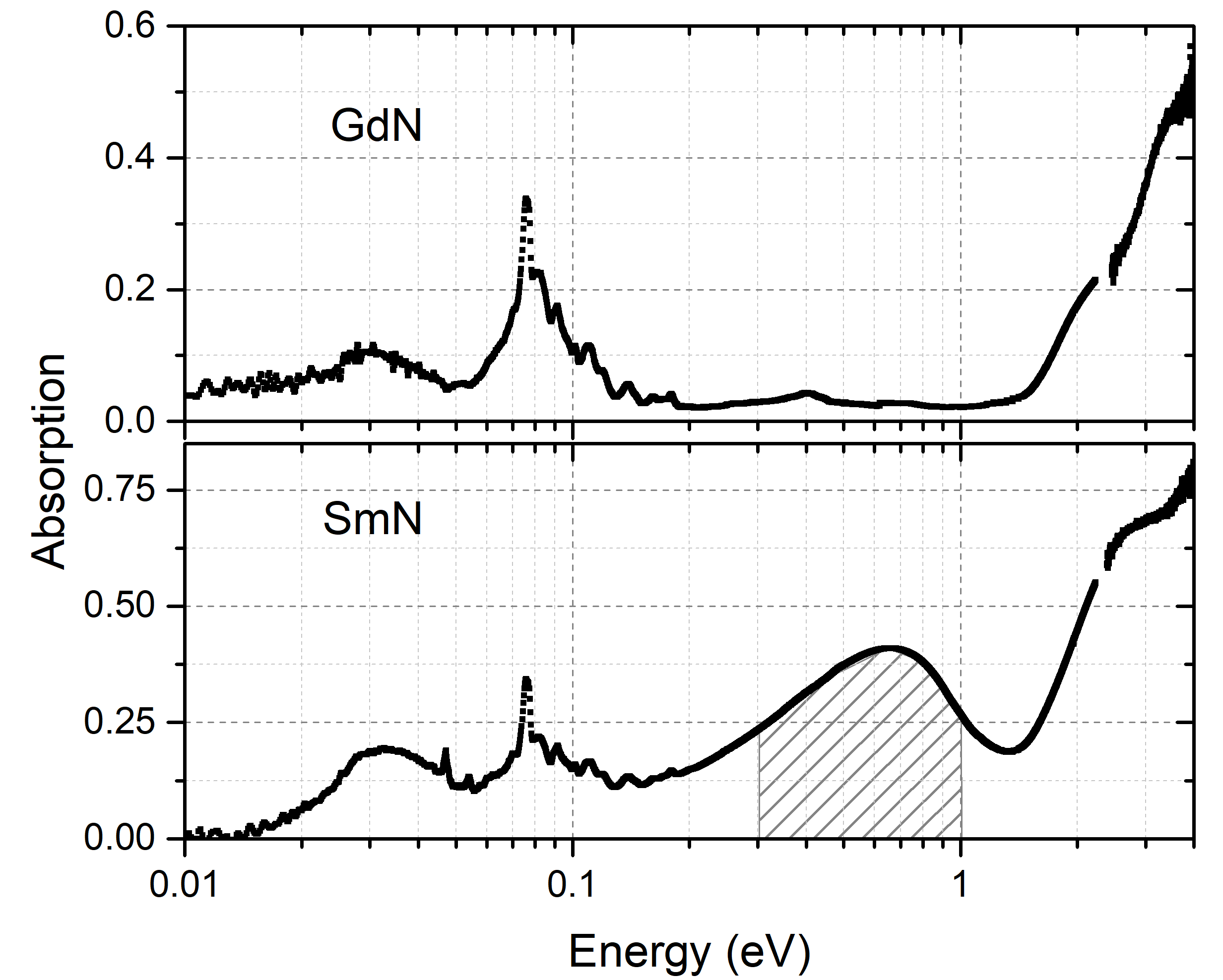}
\caption{Measurements of absorption in GdN (top) and SmN (bottom) at ambient temperature. The GdN measurement shows little to no absorption below the fundamental edge at 1.35~eV. The SmN measurement shows a similar absorption beginning above 1.2~eV but in addition a discrete feature at lower energy (identified with cross-hatching) signals additional optical transitions not present in GdN. The features near 0.1~eV shared in both plots are caused by phonon absorption in the Si substrate and AlN capping layer. The broad absorption near 0.03~eV is a phonon absorption in the LN. Differences in magnitude and oscillations at high energy are caused largely by film thickness and interference effects.}
\label{fig:1-R-T}
\end{figure}

\begin{figure}
\centering
\includegraphics[width=\linewidth]{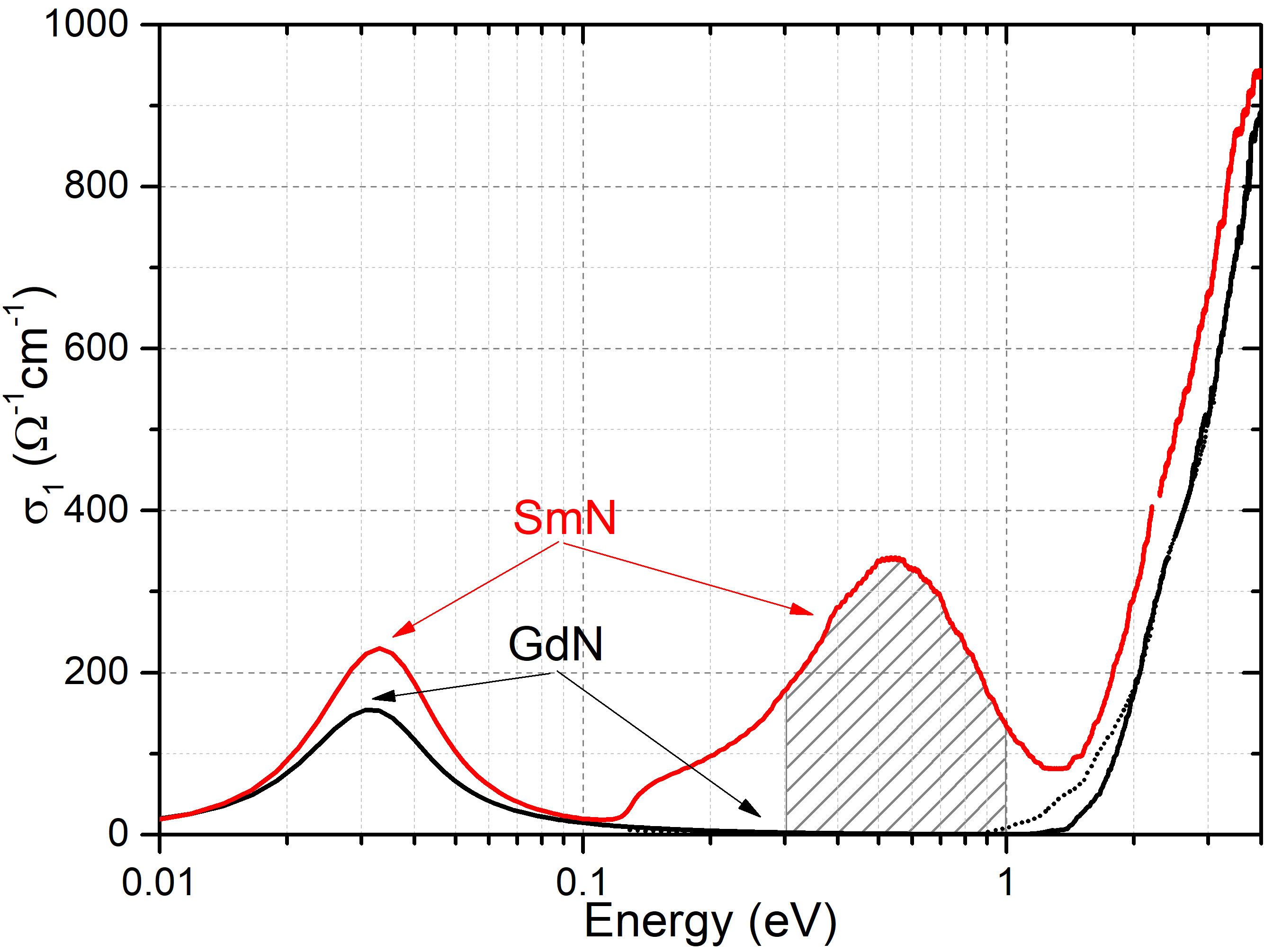}
\caption{(Colour online) The real part of the optical conductivity $\sigma_1(\omega)$ based on data visible in Figure~\ref{fig:1-R-T} for both SmN (red) and GdN (black) and GdN at 7~K (black dotted). Both materials show a strong increase in $\sigma_1(\omega)$ at high energy due to transitions from the VB to 5$d$ bands. The SmN result has an additional feature near 0.5~eV (identified with cross-hatching) which indicates transitions from the VB into a localised 4$f$ band, lying below the 5$d$. At the lowest energies a phonon absorption can be seen in each material. At low temperature the ferromagnetic red-shift of the optical edge of GdN can be seen. No measurable temperature dependence was seen in measurements of SmN samples.}
\label{fig:Sigma_1}
\end{figure}

Before the full analysis of the optical conductivity is considered it is instructive to observe the absorption \mbox{(1-R-T)} in SmN and GdN measured at ambient temperature. This is presented in Figure~\ref{fig:1-R-T} which is constructed from low energy measurements on Si and high energy measurements on sapphire substrates. Above $\sim$~1.2~eV GdN and SmN show a very similar absorption, with differences in magnitude and structure largely due to film thickness and interference effects. GdN shows very little absorption below 1.3~eV in the transparent region below the minimum optical band-gap. SmN shows a strong absorption at these energies (identified with cross-hatching). The most logical explanation for this additional absorption, based upon the electronic structure of these materials, is excitation from the VB into the unfilled majority spin 4$f$ band of SmN, not present in GdN. This is now discussed in detail, via the optical conductivity, along with low energy and temperature dependant measurements. 


The optical conductivities $\sigma_1(\omega)$ for GdN and SmN are now shown in Figure~\ref{fig:Sigma_1}. This plot shows similar features to the absorption in Figure~\ref{fig:1-R-T}, but is free from interference effects and contributions from the capping layer and substrates. We begin by discussing GdN in the inter-band region where $\sigma_1(\omega)$ increases strongly above 1.3~eV. Extrapolation of the absorption above this edge leads to a band gap of 1.35~eV in the paramagnetic phase at room temperature, dropping to 0.85~eV in the ferromagnetic phase at 7~K (7~K data are shown as the dotted series in Figure~\ref{fig:Sigma_1}). Below $\sim$~1~eV $\sigma_1(\omega)$ falls to zero below the minimum optical band-gap. 

Turning now to the SmN sample, $\sigma_1(\omega)$ above 1.2~eV is very similar to that of GdN with transposition to lower energy on the order of 0.3~eV. Again extrapolation gives a direct gap for this transition of 1.27~eV in the paramagnetic phase although this is difficult to determine given the overlap with lower energy feature. We move now to the mid / near- infra-red region spanning 0.1~eV to 1~eV. We see no absorption in the GdN below the fundamental edge. Additional absorption is now clearly visible in SmN as illustrated in Figure~\ref{fig:Sigma_1} by a broad peak in $\sigma_1(\omega)$ centred around 0.5~eV (cross-hatched as in Figure~\ref{fig:1-R-T}). This feature indicates not only absorption at lower energies than GdN but its form reveals information regarding the nature of the band, or density of states, at the bottom of the conduction band.

\begin{figure}
\centering
\includegraphics[width=\linewidth]{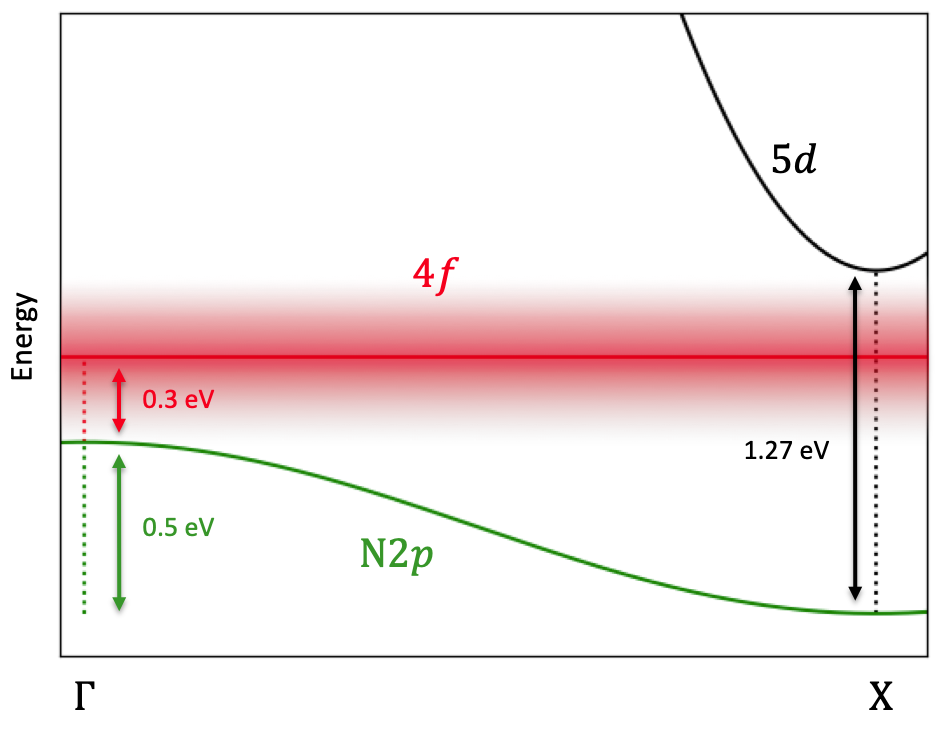}
\caption{(Colour online) Schematic band structure for SmN which is consistent with the data in Figure~\ref{fig:Sigma_1}. The Sm5$d$ band, similar to that of GdN is shown reaching a minimum at the X point, the N2$p$ valance band has its maximum at the $\Gamma$ point. The Sm4$f$ band, located via measurements shown in Figure~\ref{fig:Sigma_1}, is found beneath the 5$d$ and is shown here as dispersion-less. Shading is used to represent this as a localised band which lacks a well defined $\epsilon(k)$ relationship.}
\label{fig:BS}
\end{figure}

The task is now to construct a schematic band structure based on the measurements in Figure~\ref{fig:Sigma_1} and in the context of the calculated band structure of GdN \cite{Larson2007,Cheiwchanchamnangij2014,Mitra2008} which is largely consistent with experiment \cite{Trodahl2007,Yoshitomi2011}. The resulting schematic band structure of SmN is shown in Figure~\ref{fig:BS}. To begin the VB must be considered, this is formed from the N2$p$ states as is the case across the semiconducting members of the LN series \cite{Larson2007} with the VB maximum at $\Gamma$ and $\sim$~0.5~eV of dispersion before a stationary point at X. The filled 4$f$ states, similar to GdN, are calculated to be a minimum of $\sim$~5eV below the CB so are of little influence \cite{Larson2007}. The 5$d$ band which forms the CB minimum in many of the LN, is also qualitatively similar across the series. The data in Figure~\ref{fig:Sigma_1} show such similarity above 1.2~eV that we propose the absorption here is due to the same 2$p$ to 5$d$ transition at X that has been so well established in GdN \cite{Trodahl2007,Preston2010a,Yoshitomi2011,Larson2007}. The feature near 0.5~eV, not present in the GdN data, must then be described by excitation into some state not present in GdN. The most natural description is then the unfilled majority spin 4$f$ band of SmN implied by both anomalous Hall effect \cite{Holmes-Hewett2018} and superconductivity \cite{Anton2016a} measurements to be near the CB minimum. The separation of this feature in $\sigma_1(\omega)$ from the absorption at higher energies indicates that optical transitions begin at some point closer to $\Gamma$, as the energy of states in the VB increases when moving from X towards the VB maximum at $\Gamma$ while decreasing monotonicity moving from X to W or from X to K. 




Figure~\ref{fig:BS} shows a schematic representation of the band structure of SmN based on measurements in Figure~\ref{fig:Sigma_1}. The 4$f$ band is shown with a range of energies for any given wave-vector $k$ (represented by the shading surrounding the band in the figure) due to its highly localised nature, especially in the case of minimal hybridisation. Figure~\ref{fig:BS} additionally shows the 4$f$ band as dispersion-less, the width of the 0.5~eV feature in Figure~\ref{fig:Sigma_1} is then the width of the valance band between $\Gamma$ and X $\sim$~0.5~eV as qualitatively expected from the other LN \cite{Larson2007}. Extrapolation from the low energy side of the feature then gives the energy of this band. Figure~\ref{fig:BS} shows the 1.27~eV gap between the N2$p$ valance band and Sm5$d$ at X in black. The $\sim$~0.3~eV minimum optical gap is shown, in the limit of a dispersion-less 4$f$ band, at the $\Gamma$ point.

Given the similarities to GdN \cite{Trodahl2007,Yoshitomi2011} one may expect a temperature dependant red-shift of the optical edge in SmN. To investigate this temperature dependant measurements of both reflection and transmission were completed at various temperatures above and below the Curie temperatures of both SmN and GdN. The GdN sample showed a red-shift of the optical edge from 1.35~eV in the paramagnetic phase at ambient temperature to 0.85~eV in the ferromagnetic phase at 7~K. The optical conductivity derived from these measurements is visible as the dotted series in Figure~\ref{fig:Sigma_1}. The measured para- and ferro-magnetic gaps in GdN are consistent with previous experimental results \cite{Trodahl2007,Yoshitomi2011,Vidyasagar2014} and calculations \cite{Larson2007} which predict a 0.4~eV red-shift. Temperature dependant reflection and transmission measurements on SmN samples showed no change, thus are not shown on Figure~\ref{fig:Sigma_1}. A lack of red-shift in the 1.2~eV transition at X is perhaps not so surprising as any shift would be obscured by the additional absorption at lower energy, this can be seen by considering the magnitude of the red-shift of the GdN edge in Figure~\ref{fig:Sigma_1} and the location of the additional absorption in SmN. The lack of temperature dependence of the 0.5~eV feature is somewhat more surprising, the exchange interaction between the 4$f$ electrons below the Curie temperature appears to have very little effect on the optical absorption here.

Previous optical measurements on SmN \cite{Azeem2018} and NdN \cite{Anton2016c} films have shown an increase in absorption near 0.5~eV, similar to measurements in the present manuscript. These measurements were however not conducted to low enough energy to observe the peak of this feature and as such attributed the increase in absorption to free carriers in conductive films.


Finally the measurements at the lowest energies in Figure~\ref{fig:Sigma_1} show a phonon absorption in both GdN and SmN at 250~cm$^{-1}$ (31~meV) and 265~cm$^{-1}$ (33~meV) respectively. The measured values are $\sim$~20\% lower than those predicted \cite{Granville2009} as is seen also in DyN \cite{Azeem2013a} and in Raman measurements of several of the LN \cite{Granville2009}. As energy approaches zero and the optical conductivity approaches the DC there is no obvious contribution to $\sigma_1(\omega)$ indicative of absorption from free carriers, this indicates insulating films, consistent with transport measurements and previous experimental results indicating a semiconducting ground state \cite{Natali2013,Meyer2008,Anton2013,Anton2016a,Holmes-Hewett2018}. 


\section{Summary}

Optical measurements of reflection and transmission were undertaken on SmN and GdN thin films from ambient to cryogenic temperatures. Measurements on GdN films were consistent with previous reports \cite{Trodahl2007,Yoshitomi2011,Vidyasagar2014} yielding a band gap of 1.35~eV in the paramagnetic phase reducing to 0.85~eV in the ferromagnetic phase. Measurements on SmN samples above 1.2~eV were very similar to GdN indicating a transition at X between the VB and Sm5$d$ CB of 1.27~eV. Measurements at lower energies showed additional absorption, not present in GdN, thus indicate a low lying localised 4$f$ band, present in the conduction band of SmN but absent in GdN. A schematic band structure diagram of SmN was then created, to be consistent with the present measurements, showing the 4$f$ band sitting $\sim$~0.3~eV above the VB maximum at $\Gamma$, in the limit of a dispersion-less 4$f$ band. Neither optical edge measured in SmN showed any appreciable shift when cooled below the magnetic transition temperature. In addition measurements down to 0.01~eV relieved phonon absorption in both SmN and GdN at 250~cm$^{-1}$ (31~meV) and 265~cm$^{-1}$ (33~meV) respectively, consistent with theory \cite{Granville2009} and optical measurements on DyN \cite{Azeem2013a}. Measurements at the lowest energies showed no indication of free carrier absorption.

The identification of a localised heavy fermion conduction band in SmN provides evidence that the LN series likely harbours several more simply-structured strongly correlated materials of great interest to both theorists and experimentalists alike.

\section{ACKNOWLEDGMENTS}

The research described in the present paper was supported
by the New Zealand Marsden Fund (Grants No. 13-VUW1309 and No. 08-VUW-1309). The MacDiarmid Institute
is supported under the New Zealand Centres of Research
Excellence Programme.

\bibliography{../Master}%

\end{document}